\documentclass{emulateapj}
\usepackage{apjfonts,amsmath}

\shorttitle{The ALFALFA HI Absorption Pilot Survey}
\shortauthors{Darling, Macdonald, Haynes, \& Giovanelli}

\begin{document}
\title{The ALFALFA HI Absorption Pilot Survey:  
A Wide-Area Blind Damped Lyman Alpha System Survey of the Local Universe }
\author{Jeremy Darling\altaffilmark{1,2}, 
Erin P. Macdonald\altaffilmark{1,3}, 
Martha P. Haynes\altaffilmark{4,5}, \&
Riccardo Giovanelli\altaffilmark{4,5}}
\altaffiltext{1}{Center for Astrophysics and Space Astronomy,
Department of Astrophysical and Planetary Sciences,
University of Colorado, 389 UCB, Boulder, CO 80309-0389; jdarling@colorado.edu.}
\altaffiltext{2}{NASA Lunar Science Institute, NASA Ames Research Center, Moffett Field, CA.}
\altaffiltext{3}{Institute for Gravitational Research, School of Physics and Astronomy, Kelvin Building, University of Glasgow, Glasgow, UK G12 8QQ; e.macdonald@physics.gla.ac.uk.}
\altaffiltext{4}{Center for Radiophysics and Space Research, Cornell University, Ithaca, 
NY 14853, USA; haynes@astro.cornell.edu, riccardo@astro.cornell.edu.}
\altaffiltext{5}{National Astronomy and Ionosphere Center, Cornell University, Ithaca, NY 
14853, USA. The National Astronomy and Ionosphere Center is operated by 
Cornell University under a cooperative agreement with the National Science 
Foundation. }

\begin{abstract}
We present the results of a pilot survey for neutral hydrogen (HI) 21~cm
absorption in the Arecibo Legacy Fast Arecibo L-Band Feed Array (ALFALFA)
Survey.  This project is a wide-area ``blind'' search for HI absorption in the local universe, 
spanning $-650 ~{\rm km~s}^{-1}< cz < 17,500 ~{\rm km~s}^{-1}$ and
covering 517.0 deg$^2$ (7\% of the full ALFALFA survey).
The survey is sensitive to HI absorption lines stronger than 7.7~mJy (8983 radio sources)
and is 90\% complete for lines stronger than 11.0~mJy (7296 sources).  The total redshift 
interval sensitive to all damped Ly$\alpha$ (DLA) systems 
($N_{HI} \geq 2\times10^{20}$~cm$^{-2}$)
is $\Delta z = 7.0$ (129 objects, assuming $T_s = 100$~K and covering
fraction unity); 
for super-DLAs ($N_{HI} \geq 2\times10^{21}$~cm$^{-2}$) it is 
$\Delta z= 128.2$ (2353 objects).
We re-detect the intrinsic HI absorption line in UGC~6081 but detect no 
intervening absorption line systems.
We compute a 95\% confidence upper limit on the column density frequency distribution 
function $f(N_{HI},X)$ spanning four orders of magnitude in column
density,
$10^{19}\ (T_s/100~\rm{K})\,(1/f)$~cm$^{-2} < N_{HI} < 10^{23}\ (T_s/100~\rm{K})\,(1/f)$~cm$^{-2}$, 
that is consistent with previous redshifted optical damped Ly$\alpha$
surveys and the aggregate HI 21~cm emission in the local universe.  
The detection rate is in agreement with extant observations.
This pilot survey suggests that an absorption line search of the 
complete ALFALFA survey --- or any higher redshift, larger bandwidth, or more sensitive
survey, such as those planned for Square Kilometer Array
pathfinders or a low frequency lunar array ---
will either make numerous detections or will set a strong statistical lower limit on
the typical spin temperature of neutral hydrogen gas.
\end{abstract}

\section{Introduction}
The neutral hydrogen (HI) 21~cm (1420.405752~MHz) spin-flip transition has long been employed as an emission line tracer of 
the neutral gas content, surface density profile, kinematics, and dark matter halos of spiral galaxy
disks in the local universe.  HI 21~cm absorption, however, is rarely observed at low redshift 
outside of the Galaxy because it relies on the chance alignment of a bright radio source with foreground gas,
and the covering fraction of cold neutral gas is low.   
The very slow Einstein rate coefficient for spontaneous emission, $A_{21\rm{cm}} = 2.87 \times 10^{-15}$~s$^{-1}$,
means that a detectable 21~cm line requires either a large gas mass for emission or a high
column density for absorption; the latter operationally limits HI 21~cm absorption line studies
to damped systems ($N_{HI} > 2\times 10^{20}$~cm$^{-2}$).
But the comoving density of Damped Ly$\alpha$ (DLA) absorption systems
is only $dN/dz = 0.045\pm0.006$ per unit redshift at $z=0$ \citep{zwaan05}, 
so identifying such systems in the local universe ($z\lesssim0.1$) 
requires examination of many hundreds of sightlines.  If they can be found, HI 21~cm absorption lines 
complement emission lines;  while emission line surveys are flux-limited (mass-limited at a fixed distance), 
absorption line surveys are column density-limited (at fixed illuminating continuum), independent of 
distance, and can identify very low-mass collections of neutral gas.  HI 21~cm absorption line surveys
can potentially identify primordial atomic gas, infalling or out-going gas in the vicinity of galaxies, 
or spiral disks themselves in a distance- and dust extinction-independent fashion.  

The Arecibo\footnote{The Arecibo Observatory is part of the National Astronomy and Ionosphere Center, which is operated by Cornell University under a cooperative agreement with the National Science Foundation.} Legacy Fast Arecibo L-Band Feed Array (ALFALFA) Survey is a 7000 deg$^2$
extragalactic HI 21 cm emission line survey.  Compared to the HI Parkes All Sky Survey 
\citep[HIPASS;][]{hipass04}, the 
ALFALFA survey is 8 times more sensitive, has 4 times the angular resolution, 3 times the 
spectral resolution, and a factor of 1.4 more bandwidth \citep{giovanelli05}, 
and provides the first data set approaching the sensitivity and areal coverage
required for a truly ``blind'' HI 21 cm absorption line search.  Such a search is effectively a search 
for DLAs that does not rely on UV-bright quasar sightlines (and thus
can detect dusty DLAs).  

\begin{figure*}[ht]
\plotone{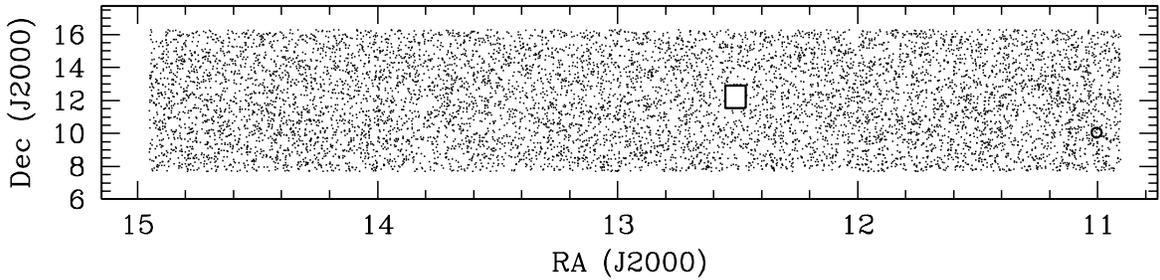}
\caption{
Radio sources searched for HI absorption in the ALFALFA survey.
The blank boxed region was not searched due to interference from 
the strong radio source in M~87.  
The circled source is the re-detected intrinsic HI absorption line in UGC~6081
(\S \ref{subsec:UGC6081}).
\label{fig:sources}}
\end{figure*}
 
In this paper, we describe the ALFALFA HI 21 cm absorption pilot survey, a test of the notion 
that damped Ly$\alpha$ systems can be detected ``blindly'' in the local Universe in a shallow, 
large area spectroscopic survey for HI emission from galaxies.  We present the methods and 
results of our survey, along with follow-up observations conducted at the Green Bank Telescope\footnote{The National Radio Astronomy Observatory is a facility of the National Science Foundation operated under cooperative agreement by Associated Universities, Inc.}  (GBT),
and demonstrate that it is possible to place meaningful upper limits on the column density frequency 
distribution function.
% as well as on $\Omega_{HI}$ at $z\sim0$.   
The pilot survey results suggest that 
a search of the full 7000 deg$^2$ of the ALFALFA survey for HI 21~cm absorption would be 
worthwhile, both for understanding cold neutral gas in the local Universe and in preparation
for planned absorption line surveys at high and low redshift with upcoming new radio telescopic facilities.

\section{Searching for Absorption Lines in the ALFALFA Survey}\label{sec:search}

%[describe the data]\\
The ALFALFA survey offers the combined sensitivity, areal coverage, and redshift span to make
a ``blind'' HI 21~cm absorption line search feasible and astrophysically interesting.  
The complete ALFALFA survey will cover $\sim$7000 deg$^2$ and is expected to detect 
more than 25,000 galaxies in HI 21~cm emission \citep{giovanelli05}. 
The data 
products of the survey are dual-polarization 2.4$^\circ \times$2.4$^\circ$ 1024-channel spectral
data cubes spanning 25 MHz each.  There are four spectrally overlapping cubes per sky position spanning the range $-2000 <  cz < 17,900$~km~s$^{-1}$ \citep{giovanelli05}. 
The rms noise per spectral channel after smoothing to 10~km~s$^{-1}$ is 2.2~mJy \citep{saintonge07}.

%[describe the pilot survey]\\
Our ALFALFA HI absorption pilot survey examines a 517 deg$^2$ subset 
of the mature ALFALFA data product in the northern Virgo Cluster region, spanning 
$10.9^h < \alpha < 14.95^h$ and $+7.7^\circ < \delta < +16.3^\circ$ (122 spectral cubes).
We omit a 1.6 deg$^2$ region centered on M~87; the strong 
radio source in M~87 (220 Jy at 1.4 GHz) creates spectral standing 
waves that render this region inaccessible to weak line spectroscopy with Arecibo \citep{giovanelli07}.  

Figure \ref{fig:sources} shows the survey area and distribution of 
radio sources with $8.4~{\rm mJy} < S(1.4~{\rm GHz}) < 5.3$~Jy.
%(see below for an explanation of these flux density limits).  
We searched a velocity range 
$-650~{\rm km~s}^{-1}< cz < 17,500~{\rm km~s}^{-1}$, with notable 
gaps due to radio frequency interference or Galactic HI emission
at $cz=0$, 2200, 3300, and 8800~km~s$^{-1}$ of about 200~km~s$^{-1}$ width
plus a large gap at 15,000--16,000~km~s$^{-1}$  \citep{giovanelli07}.
The gaps in the spectral coverage 
render $\sim$10$\%$ of the velocity range unsearchable. % (RG07 claim 15\%).  
The net search covers $\Delta z = 0.054$ along each line of sight 
plus the range $-650~{\rm km~s}^{-1}< cz < -100~{\rm km~s}^{-1}$.
This translates into a comoving path length of  
$\Delta X = 0.0571$\footnote{%[$\Delta z = 0.0545$ is $\Delta X = 0.0571$.  ]
%(does it makes sense for dX to be greater 
%than dz (as opposed to less)? Yes.]
We assume a $\Omega_\Lambda = 0.73$ and $\Omega_m = 0.27$ cosmology
\citep{wmap7}.  
% (WMAP7 values)
Making no assumptions about global curvature, 
$\Omega_k = 1-\Omega_\Lambda - \Omega_m$, the comoving path length is 
$\Delta X = \int_{z_1}^{z_2}dz
(1+z)^2/\sqrt{(1+z)^2(1+z\Omega_m)-z(z+2)\Omega_\Lambda}$.
We use the $\Delta X$ quantity only in the calculation of the column
density frequency distribution function (Sec. \ref{sec:analysis});
otherwise, $\Delta z$ is used to describe the survey.}. 

%[describe the search methods]\\
Our line search relies on the optimized matched filter detection methods developed 
by \citet{saintonge07}:  we multiply the continuum-subtracted data cubes by $-$1 and run
the line-finding algorithm as if we were searching for narrow Gaussian HI emission lines.  
\citet{saintonge07} obtains matched filter (not single-channel) signal-to-noise  (S/N) thresholds for both reliability
(detections are reliably not spurious) and completeness (all lines are detected) of
5.5 and 6.5, respectively, for narrow lines in the ALFALFA survey.  
Our absorption line search used the conventional ALFALFA
matched filter S/N of 4.6 to search for lines with central depth greater than 
7.7 mJy toward continuum\footnote{All radio continua used in the selection of sources and the analysis of 
survey results were obtained from the NRAO VLA Sky Survey (NVSS) catalog and represent integrated 
flux densities \citep{condon98}.}
sources with flux densities greater than 8.4 mJy (this continuum threshold corresponds to 
a line center optical depth of 2.5 and a neutral hydrogen column density of $1.4\times10^{22}$~cm$^{-2}$
assuming hydrogen spin [21~cm line excitation] temperature $T_s
=100$~K, a source covering fraction $f$ of unity, and a
30~km~s$^{-1}$ FWHM line width).  We chose this continuum threshold
to allow for the possibility of detecting very large column densities; while expected to be very rare, 
the number of sources probing such regimes on the sky is large.  Our low S/N threshold does identify 
many spurious absorption lines, which we subsequently assess using five
criteria:  (1) association with NVSS radio sources (the 45\arcsec\ NVSS beam is well-matched to the 1\arcmin\ ALFALFA pixels); 
(2) similar line properties in both linear polarizations; 
(3) unresolved angular size; (4) no association with radio frequency interference; and
(5) no association with spectral standing waves.  Possible absorption 
lines that meet these criteria were subsequently re-observed with the GBT (Section \ref{sec:obs}).

\begin{figure}
\plotone{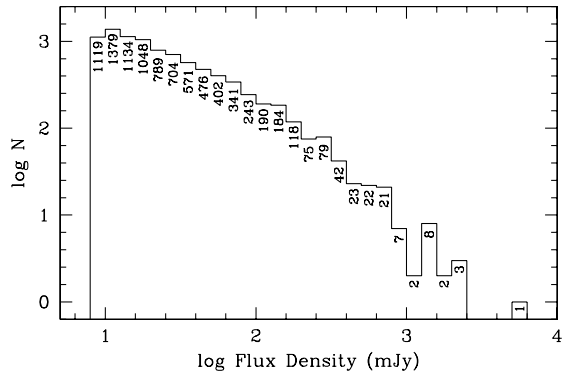}
\caption{
Flux density distribution at 1.4 GHz of the radio sources searched for HI absorption 
in the ALFALFA Survey.  Vertical numbers indicate the number of sources in each bin. \\
\label{fig:fluxes}}
\end{figure}

The vast majority of lines detected using our S/N threshold of 4.6 were obviously spurious, 
in large part because 
we were searching below the reliability threshold.  For the
calculation of column density statistics
(Section \ref{sec:analysis}), we
require confidence that detected lines are real (reliability) and that a non-detection is in fact a 
non-detection (completeness), so we impose the more severe S/N cut of 6.5, 
corresponding to the \citet{saintonge07} estimate of the 90\% completeness limit.  For the gaussian 
narrow line template with FWHM of 30~km~s$^{-1}$, this corresponds to a line peak of 11.0 mJy.  Allowing
for opaque lines, we make a continuum cut at the same level.  For the full line search, there are
8983 radio sources with $S(1.4~\rm{GHz})>8.4$~mJy, but for the restricted column density statistics, 
we reduced the sample to 7296 sources with $S(1.4~\rm{GHz})>11.0$~mJy.  The distribution of continua
for the full sample is shown in Figure \ref{fig:fluxes}.

%[strong sources and standing waves; describe the sample of strong sources and point to next sections]\\
Supplemental observations were required for two samples:  candidate HI absorption systems, and 
strong continuum sources.  Strong sources create spectral standing waves between the primary reflector and the 
superstructure blocking the aperture of Arecibo, frustrating the detection of weak lines.  
It is possible that a technique can someday be developed which replicates the ``double position-switching''
mode of Arecibo in software using continuum templates within the survey data.  But in the meantime, 
it was necessary to observe strong continuum 
sources, $S(1.4~\rm{GHz})>220$~mJy, individually (Section \ref{sec:obs}).
The strong sources in our sample span the range 220~mJy$< S(1.4~\rm{GHz})<5.3$~Jy 
(and exclude M~87 at 220~Jy), but the majority fall in the range 220--500~mJy.

\section{Observations and Data Reduction}\label{sec:obs}

We observed 13 candidate absorption systems and 
250 ``strong'' sources (NVSS flux density greater than 220 mJy) with the GBT 
in April through August, 2008.  The GBT is particularly well-suited to weak spectral line detection toward 
strong continuum sources thanks to its unblocked aperture.  
Observations of all sources were designed to mimic the spectral resolution, spectral coverage, and 
sensitivity of the ALFALFA survey.  

For the follow-up observations of line candidates, we observed for a total of 10 minutes on-source in a 
5-minute position-switched mode using two linear polarizations, 9-level sampling, 6-second records,
and a 50 MHz bandwidth centered on the tentative line.  
The strong sources were observed in a search mode using two overlapping 50~MHz bands 
% [centered on $-$15 and $-60$ MHz from 1420.405752]
spanning 95~MHz from 1335.4 to 1430.4~MHz ($cz = -2090$ to 19,080~km~s$^{-1}$) 
for 5 minutes on-source with a reference ``Off'' spectrum recorded roughly 
every fifth source.  We used two linear polarizations, 9-level sampling, and 3-second records.
For all observations, a calibration diode signal was injected for half of each record.  

Each individual off-source-flattened and calibrated 
spectral record was manually examined and flagged for RFI, then averaged 
in time and polarization.  After polynomial baseline subtraction, we achieved 
1.3~mJy rms noise in 10.4~km~s$^{-1}$ channels (at 1400~MHz) toward the candidate absorbers
after Hanning and Gaussian smoothing and averaging polarizations.  
We reached $2.2\pm0.2$~mJy rms noise toward the 151 strong sources with continua 
in the range 220--370~mJy, $2.5\pm0.3$~mJy toward the 83 sources of 370--1000~mJy, 
$2.8\pm0.3$~mJy toward the 12 sources of 1.0--2.0~Jy, and 
$4.3\pm1.2$~mJy toward the 4 sources of 2.0--5.3~Jy (the quoted uncertainties
span the range of rms noise values in each source grouping).
% {\red (numbers:  2.2 mJy ($\pm0.2$) in 10.3 km/s channels in Groups 1--2 sources 
%(220--370 mJy); 2.5$\pm0.3$ mJy in Group 3 (370--600 mJy); xx in Group 4 ($>600$ mJy)}.  
All GBT data reduction and analysis was performed with 
GBTIDL\footnote{GBTIDL (\url{http://gbtidl.nrao.edu/}) is the data reduction package produced by NRAO 
and written in the IDL language for the reduction of GBT data.}.

%{\red [Have a look at the GBT strong sources final spectra.]}

%{\red [Is it still appropriate to use the fixed noise limit on the strong sources, or do we need to 
%vary the noise level??]}

\section{Results}

\subsection{Column Density Limits}\label{subsec:column-density}

\begin{figure}
%\epsscale{1.15}
\plotone{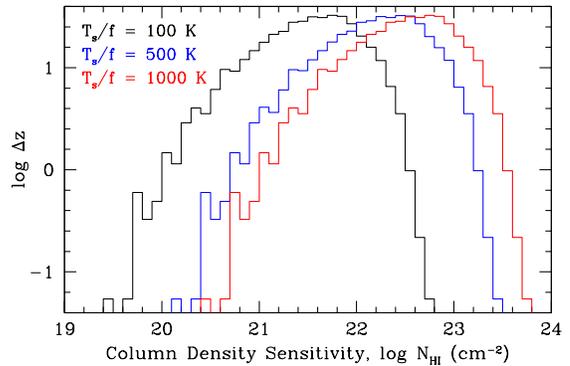}
\caption{Redshift search path versus HI column density sensitivity.  
Each column density bin indicates the lower bound on 
the column density detectable toward the sources in that bin.  
This distribution represents the reliable and complete sample
with a flux density cut of $S(1.4~{\rm GHz}) > 11$~mJy
(as opposed to the 8.4~mJy cut for the sources displayed in 
Figures \ref{fig:sources} and \ref{fig:fluxes}). %, with the exception 
%of a few of the strongest continuum sources (Section
%\ref{subsec:column-density}) .
The black, blue, and red histograms indicate the distributions
for spin temperature modulo covering factor ($T_s/f$) values of 100,
500, and 1000~K, respectively.  \\
\label{fig:columns}}
\end{figure}

We made no new detections of HI 21 cm absorption (intrinsic or
intervening) in the pilot survey:  
no ALFALFA absorption line candidates were confirmed by GBT observations, and 
no strong sources observed with the GBT show absorption lines.  Despite
the lack of new detections, the uniform data set and large number of 
continuum sources in the survey allow us to place limits on the 
column density frequency distribution function $f(N,X)$ (Section \ref{sec:analysis}).  
Each non-detection of HI absorption toward a radio continuum
source places an upper limit on the HI column density of 
\begin{equation}\label{eqn:NHI}
N(HI) < 1.8\times 10^{18}\ (T_s/f)\, 1.064\  {\rm FWHM} \cdot \tau~\rm{cm}^{-2}
\end{equation}
where the limit on optical depth is set by S/N$= 6.5$ corresponding to the completeness limit,
\begin{equation}\label{eqn:tau}
\tau < -\ln\left(1-\frac{11~{\rm mJy}}{S(1.4~{\rm GHz})}\right),
\end{equation}
and we assume the minimum FWHM matched filter template line width used for the search, 
30~km~s$^{-1}$. %, a source covering fraction of unity, and a spin temperature $T_s = 100$~K.  
For the strong sources observed independently at the GBT, we adjust the 11~mJy line limit above
according to the rms noise:  it is unchanged for the 220--370~mJy sources (2.2~mJy rms), but 
increases to 28~mJy for the strongest source at 5.3~Jy.  We can generally detect {\it any} DLA 
toward the strong sources.  Figure \ref{fig:columns}
shows the redshift path searched for each column density limit.
 
%{\red [words about number of DLA-sensitive objects, dz, and for super-DLAs (as per abstract).  Also
%mention full range of columns (sub-DLA).]}
For a source covering fraction of unity and a spin temperature $T_s =
100$~K, the total redshift 
interval sensitive to all damped Ly$\alpha$ (DLA) systems ($N_{HI} \geq 2\times10^{20}$~cm$^{-2}$)
is $\Delta z = 7.0$ (129 objects); for super-DLAs ($N_{HI} \geq 2\times10^{21}$~cm$^{-2}$) it is 
$\Delta z= 128.2$ (2353 objects).  The lowest detectable column
density under these assumptions is $3\times10^{19}$~cm$^{-2}$ (but see
Figure \ref{fig:columns}).

\subsection{Re-detection of Intrinsic HI Absorption in UGC 6081}\label{subsec:UGC6081}
\begin{figure}
\plotone{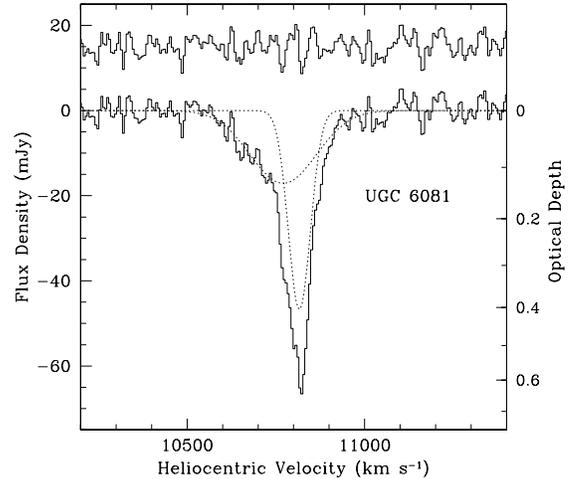}
\caption{HI absorption spectrum of UGC~6081.  The dotted lines show
the two Gaussian fits to the spectrum, and the upper spectrum shows
the residual spectrum (data$-$model) offset by +15~mJy for clarity.  
The optical depth plotted on the right assumes a continuum flux density
of 140~mJy and a covering factor of unity (see \S \ref{subsec:UGC6081}).
\label{fig:UGC6081}}
\end{figure}

We have re-detected the strong intrinsic HI absorption line in the 
interacting system
UGC~6081 at $cz \simeq  10,800$~km~s$^{-1}$, previously detected
by \citet{bothun83} and \citet{williams83}.  
Gaussian fits to the ALFALFA HI absorption spectrum (Figure \ref{fig:UGC6081})
are listed in Table 
\ref{tab:UGC6081}.  A two-component fit produces a residual spectrum 
nearly consistent with the rms noise (a three-component fit is not warranted
by the data).  

\begin{figure}
\plotone{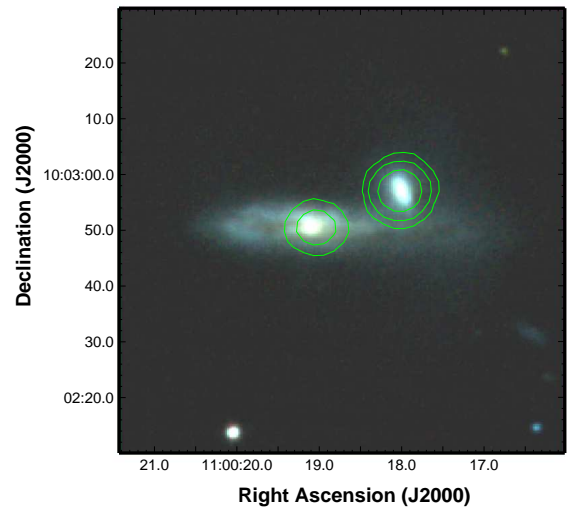}
\caption{Sloan Digital Sky Survey \citep{york00} false color image of UGC~6081 with FIRST 20 cm 
radio continuum contours (green; 3.0, 10.4, and 36.3 mJy).  
Red, green, and blue correspond to Sloan $z$, $i$, and $r$.
The ALFA beam is larger than the displayed image, but the centroid of the line is well-matched 
to the continuum sources.
\label{fig:UGC6081sdss}}
\end{figure}

UGC~6081 lies in a region not covered by IRAS, but it does appear in the 2MASS
catalog as multiple sources \citep[$K_s= 12.8$--13.5,][]{skrutskie06}
and is likely a luminous infrared galaxy (LIRG).  Despite the 
strong HI absorption, the main 18~cm OH lines were not detected in this system
at the level of 5.2~mJy rms in 9~km~s$^{-1}$ channels \citep{kazes85}.
The FIRST survey resolves this system into two radio sources:
the NW component is the stronger at 140.3(0.1)~mJy; the SE component 
has an integrated flux density of 30.8(0.1)~mJy \citep[][Figure \ref{fig:UGC6081sdss}]{white97}.  
\citet{bothun83}
and \citet{williams83} assumed that the radio continuum was a single
source and incorrectly used the aggregate flux density at 1370 or 1400~MHz to 
compute the optical depth and HI column density.  
Since no interferometric
spectral line observations have been conducted of this system to date, and
the 16\arcsec\ separation of the two continuum sources is not resolved
by the ALFALFA survey (the ALFA beam is 3.3\arcmin$\times$3.8\arcmin), 
the source(s) of the HI absorption remains somewhat ambiguous.  
Note that the measured redshifts of the two merging components
are consistent with both HI absorption lines and cannot resolve the HI
provenance \citep{geller84}.
% uncertainty in the main galaxy is 20 km/s (Arecibo) and 100 km/s in
% the NW compt (MMT) --- Geller et al 1984
The stronger HI component must
originate in the stronger (NW) component of UGC~6081 because the line
is deeper than the flux density of the SE continuum source.  
The weaker, blueshifted
HI component, however, could arise in either radio continuum source. 
We compute optical depths assuming that all HI absorption 
originates in the NW radio source, and we obtain a very large total HI column density 
of $N({\rm HI}) = 1.13(1)\times10^{22}\ (T_s/100~{\rm K})\,(1/f)$~cm$^{-2}$ 
by direct integration of the observed 
spectrum (rather than from the Gaussian fits).  This column density 
is a factor of 1.4 and 2.8 larger than the values obtained by 
\citet{williams83} and \citet{bothun83}, respectively, due in large part to 
larger and confused continua employed in the previous calculation of optical depths.
Note that if $T_s>100$~K, $f<1$, or the blue line were associated with
the SE component, then the already very large column density would be
even larger.

Our re-detection of HI absorption in UGC~6081 illustrates the power of the 
ALFALFA survey --- designed as an HI emission line survey --- to detect
HI absorption (intrinsic and intervening) in a new completely ``blind'' survey mode.  

\begin{deluxetable*}{clclccccc} 
\tabletypesize{\scriptsize}
\tablecaption{Intrinsic HI Absorption Lines Toward UGC 6081\label{tab:UGC6081}}
\tablecolumns{9}
\tablewidth{0pt}
\tablehead{
\colhead{Line} & 
\colhead{$\nu$}  & 
\colhead{$v$} &
\colhead{$z$} & 
\colhead{FWHM}  & 
\colhead{Depth}  &
\colhead{$\tau$} &
\colhead{$\int\tau\,dv$} & 
\colhead{$N_{HI}$} \\
\colhead{} & 
\colhead{(MHz)} & 
\colhead{(km s$^{-1}$)} &
\colhead{} & 
\colhead{(km s$^{-1}$)} &
\colhead{(mJy)} & 
\colhead{} & 
\colhead{(km s$^{-1}$)} &
\colhead{(cm$^{-2}$)} 
}
\startdata
1 & 1370.943(4) & 10816.3(0.9) & 0.036079(3) & 73(3)  & $-$46.7(1.6) &
0.40(2) & 31.5(1.6) &  $0.57(3)\times10^{22}\ (T_s/100~{\rm K})\,(1/f)$\\
%724.823(1) & 0.959658(4) & +193(31)  & 27(1) & $-$68(2) & 0.0176(5) & 0.50(2)   & $9.1\times10^{17}.\,(T_{\rm s}/f)$\\
2 & 1371.14(3) & 10770.8(5.8) & 0.03593(2) & 220(10)& $-$17.1(1.4) &
0.13(1) & 30.4(2.7) & $0.55(5)\times10^{22}\ (T_s/100~{\rm K})\,(1/f)$\\[3pt]
%\hline
\vspace{5pt}
Total& 1370.91(3) & 10823.1(5.5) & 0.03610(2) & 86(2) & $-$66.5(2.2) &
0.64(3) & 62.1(0.7) & $1.13(1)\times10^{22}\ (T_s/100~{\rm K})\,(1/f)$
\enddata
\tablecomments{HI absorption properties of UGC~6081.  
We list the the observed line center frequencies, velocities, and redshifts 
(barycentric, optical definition), the line widths and depths, apparent 
and integrated optical depths (see Section \ref{subsec:UGC6081}), and 
HI column density.  Values in parentheses indicate uncertainties in the
final digit(s).  The FWHM and integrated optical depth values are 
in the rest frame of UGC~6081.
The first two rows in the Table indicate measured and derived
quantities obtained from Gaussian fits to the spectrum, while
the last row in the Table lists numerical values for the entire
absorption system obtained without Gaussian fits.\\}
\end{deluxetable*}

\section{Analysis}\label{sec:analysis}

%\subsection{The Column Density Frequency Distribution Function}

%[Check X for this interval by doing the integral explicitly (numerically).  Answer:  $\Delta X = 0.0571$; 
%careful treatment of redshift ranges yields $\Delta z = 0.0545$, including a small bit of negative 
%velocity space.]

We compute the column density frequency distribution function $f(N_{HI},X)$ following \citet{cooksey10}
using the limiting column density sensitivity estimates shown in Figure \ref{fig:columns}:
\begin{equation}\label{eqn:fofN}
f(N_{HI},X)<\frac{\lambda_{max}}{\Delta N_{HI} \Delta X}
\end{equation}
where $\lambda_{\rm max}$ is the Poisson upper limit on the detection rate of absorption systems with 
column density $N_{HI}$  in interval $\Delta N_{HI}$ and $\Delta X$ is the comoving absorption
path length searched.  In this case, the search interval is uniform for all sources, so 
$\Delta X = \delta X\,n_{\rm sens}$, where $\delta X=0.0571$ is the corrected redshift interval 
$\delta z=0.0545$, and $n_{\rm sens}$ is the number of sources toward which observations were 
sensitive enough to detect a column density equal to or greater than $N_{HI}$ (Figure \ref{fig:columns}
shows the nearly identical quantity $\Delta z = \delta z\,n_{\rm sens}$ versus the column density sensitivity).
A 95\% confidence upper limit on the Poisson rate is $\lambda_{\rm max}= 3.00$ when no detections 
are made. %and $\lambda_{\rm max}= 4.64$ for a single detection (UGC~6081).  

\begin{figure}
%\epsscale{1.15}
\plotone{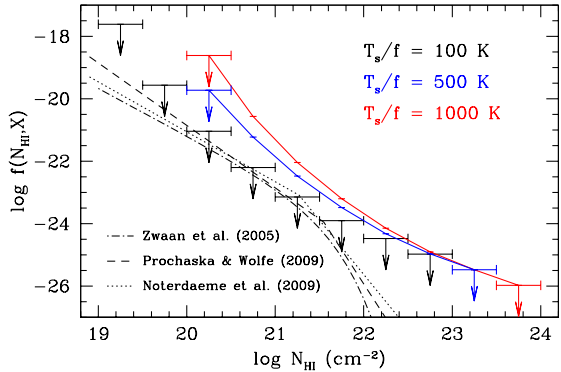}
\caption{
The column density frequency distribution function, $f(N_{HI},X)$.
All points are 95\% confidence upper limits
and the horizontal bars indicate the column density range spanned by
each limit (some arrows and bars are omitted for clarity).
The long-dashed line indicates the broken power law  
fit by \citet{proch09} for absorption systems at $z=2$--4, and the
dotted line indicates a revised fit by \citet{noterdaeme09}.
The dash-dot line shows the gamma distribution fit for
DLAs in $z=0$ galaxies mapped in HI 21~cm emission 
by \citet{zwaan05}.
The black, blue, and red upper limits assume spin temperatures
modulo covering factor ($T_s/f$) of 100,
500, and 1000~K, respectively.  The effect of increasing spin
temperature is to shift the locus down and to the right because the 
column density increases linearly with $T_s$ (Eqn.\ \ref{eqn:NHI}).
\label{fig:fofN}}
\end{figure}

\begin{figure}
%\epsscale{1.15}
 \plotone{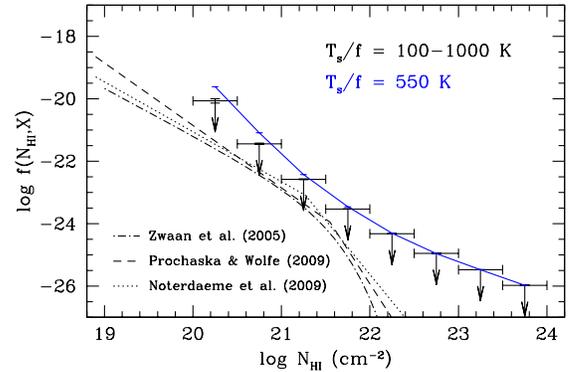}
\caption{
The column density frequency distribution function, $f(N_{HI},X)$,
calculated from a uniform random distribution of spin temperatures 
(modulo covering factor) in the
range $T_s/f=100$--1000~K.  All points are 95\% confidence upper limits
and the horizontal bars indicate the column density range spanned by
each limit (some arrows and bars are omitted for clarity).
The vertical error bars indicate the 1~$\sigma$ range in the 95\% 
confidence upper limit obtained from 1000 trials.  
For reference, the blue line indicates the upper limit obtained for
a uniform spin temperature of 550~K (the mean of the distribution).
The long-dashed line indicates the broken power law  
fit by \citet{proch09} for absorption systems at $z=2$--4, and the
dotted line indicates a revised fit by \citet{noterdaeme09}.
The dash-dot line shows the gamma distribution fit for
DLAs in $z=0$ galaxies mapped in HI 21~cm emission 
by \citet{zwaan05}.
\label{fig:fofNrandom}}
\end{figure}

Figure \ref{fig:fofN} shows the column 
density frequency distribution 95\% confidence limits calculated for a
range of $T_s/f$ values, but in each case a single value is assumed for
the entire survey.  
To address this, we also calculated the 95\% confidence limits 
on $f(N_{HI},X)$ for 1000 trials in which $T_s/f$ is uniformly
randomly distributed for each sight line in the range 100--1000~K.
Figure \ref{fig:fofNrandom} compares the result of a randomly
distributed $T_s/f$ to the mean of the distribution (550~K), showing
little difference between the two distributions except in the lowest
column density bins.  This demonstrates
that despite the unconstrained nature of the spin temperature and
covering factor, a distribution of values is well-represented in 
the $f(N_{HI},X)$ upper limit locus by a mean value.  Also, although $T_s/f$ is
unconstrained source-by-source, the prior on this quantity can
still produce a useful constraint on $f(N_{HI},X)$, increasingly so as
surveys get larger or more sensitive (Section \ref{sec:discussion}).
Figures \ref{fig:fofN} and \ref{fig:fofNrandom} compare our column
density frequency distribution limits to measured $f(N_{HI},X)$ distributions from
the \citet{zwaan05} $z\simeq0$ HI 21~cm emission line analysis
and from the optical surveys of redshifted DLAs at $z=2$--4 by \citet{proch09} and
\citet{noterdaeme09} and show them to be consistent in all cases.

%{\red [discussion of effect of bin width choice on the distribution limits]}\\
The calculation of $f(N_{HI},X)$ is influenced by the choice of column density interval
$\Delta N_{HI}$ because we are calculating upper limits:  the numerator in Equation \ref{eqn:fofN}
is independent of the number of sources whereas the denominator grows with $\Delta N_{HI}$.
We chose $\Delta N_{HI}=0.5$~dex as a compromise between good statistics in most 
bins and resolution of $f(N_{HI},X)$ along the column density axis, but it should be borne in mind
that these limits do depend on this choice.  With coarser sampling, $f(N_{HI},X)$ moves lower but
does not become inconsistent with the fits to $f(N_{HI},X)$ obtained
by \citet{zwaan05}, \citet{proch09}, and \citet{noterdaeme09}.

The column density sensitivity 
is proportional to the combination of factors FWHM$\cdot T_s/f$;
%($f$ here is the source covering fraction); 
if this quantity decreases, 
the sources are redistributed into lower column bins, driving the $f(N_{HI},X)$ values lower.  
A line FWHM smaller than the assumed 30~km~s$^{-1}$ is likely in many DLAs, 
suggesting that an average value of $T_s/f=100$~K is probably too low
(Figure \ref{fig:fofN}).

A source of systematic uncertainty in our calculation of $f(N_{HI},X)$ is the unknown redshift 
distribution of the illuminating radio sources.  We have assumed for this analysis that all radio sources lie
at $cz>17,500$~km~s$^{-1}$, but there are certainly cases to the contrary (e.g., UGC~6081).  
This error operates in one direction:  it diminishes the total redshift search path of the survey, driving 
our $f(N_{HI},X)$ limit upward.  
%But the redshift distribution of radio sources is not necessarily independent of 
%continuum flux density and therefore not independent of column density sensitivity, so the effect of 
%unknown redshifts may also alter the shape of $f(N_{HI},X)$ somewhat.  
%Given the large number of 
%continuum sources {\red [and etc...]}, we estimate the size of the effect on $f(N_{HI},X)$ to be no
%more than {\red xx} dex.
The size of this effect, however, is very small:  using the redshift distribution of NVSS sources
with $S(1.4~{\rm GHz}) > 10$~mJy determined by \citet{brookes08} and fit by \citet{dezotti10}, 
we calculate that no more than about 60 sources in the 517~deg$^2$ pilot survey lie within the 
observed redshift range, which is 0.8\% of the sample.  The redshift distribution of these sources will also 
be skewed toward the high redshift side of the survey, so the effect of this incomplete redshift
coverage is truly negligible.  
%These statistics suggest that the case of HI 21 cm absorption in
%UGC~6081 is all the more remarkable for its apparent rarity.

%{\red [Add the possibility that assumed FWHM is incorrect or that covering factor is not unity, and 
% trace the effect of these on the distribution and interpretation.]}  

%\subsection{Moments of the Column Density Frequency Distribution Function}

%{\red [Discussion of why we don't do moments, including Omega.]}\\
Since we obtain only upper limits on the column density frequency distribution $f(N_{HI},X)$, its 
moments, such as the HI mass density ($\rho_{HI}$ or equivalently $\Omega_{HI}$) and the comoving
covering fraction $\ell(X)$, are dominated by the endpoint limits 
and thus not well constrained by our observations.  
It is likely that such moments can be computed from future surveys (Section \ref{subsec:future}).
% {\red Chuck the intrinsic stuff b/c it is not intervening...}
%  We can, however, estimate $\Omega_{HI}$ directly from our {\it single} detection:
% \begin{equation}\label{eqn:Omega}
% \Omega_{HI} = {H_\circ\over c} {m_H\over\rho_{crit}}{\sum_i N_{HI,i}\over\Delta X} 
% \end{equation}
% \citep{lanzetta91}\footnote{We assume $H_\circ=71$~km~s$^{-1}$~Mpc$^{-1}$.}.
%  Including only the $\Delta X$ for which the line toward UGC~6081 could be detected (6412 
%  sources), we obtain $\Omega_{HI,\circ} = 4.2\times10^{-4}\ (T_s/100~\rm{K})\ (1/f)$, 
%  which is consistent with previous $z\sim0$ HI 21~cm emission line measurements
%  (\citet{zwaan05a} obtain $\Omega_{HI,\circ} = 3.5\pm0.4\pm0.4\times10^{-4}\ h_{75}^{-1}$)
%  and low redshift DLA samples 
%  (\citet{rao06} obtain $\Omega_{HI} = 7.5\pm2.8\times10^{-4}\ h_{70}^{-1}$ for $0.11<z\leq0.90$\footnote{We have removed the He mass correction $\mu=1.3$ to convert $\Omega_{\rm DLA}$ 
%  into $\Omega_{HI}$.}),
%  given the $\sim$100\% uncertainty stemming from a sample of one.  The equivalent 
%  mass density is $\rho_{HI} = 0.58\times10^8 \ (T_s/100~\rm{K})\,(1/f)$~M$_\odot$~Mpc$^{-3}$, 
%  consistent with the lowest redshift measurement of \citet{proch09}:  
% $\rho_{HI} = 0.555^{+0.095}_{-0.096}\times10^8$~M$_\odot$~Mpc$^{-3}$ for $z=[2.2,2.4]$.
% But the larger point here is that 
Our lack of detections is perfectly consistent with expectations and recent work on damped 
Ly$\alpha$ lines and HI 21~cm emission (Section \ref{sec:discussion}).

\section{Discussion}\label{sec:discussion}
Our upper limits on the column density frequency distribution function are consistent with previous 
determinations of $f(N_{HI},X)$, both at high redshift 
\citep[][via Ly$\alpha$ absorption]{proch09,noterdaeme09} and low \citep[][via HI 21~cm emission]{zwaan05}.
Figures \ref{fig:fofN}  and \ref{fig:fofNrandom} show that an expanded
survey would either start making absorption line detections or 
will place a strong statistical limit on the hydrogen spin temperature 
to source covering fraction ratio ($T_s/f$).
We suspect that it will be the latter, but in any case, our results provide 
strong motivation for a more sensitive or a larger area survey. 
For example, a search for absorption lines in the full ALFALFA survey would
increase the comoving search path $\Delta X$ by a factor of 14 (1.15 dex) in each $N_{HI}$
bin, which will place the $f(N_{HI},X)$ 95\% confidence limits in conflict with
previous measurements of the distribution over the range 
$10^{20}$~cm$^{-2}\lesssim N_{HI}\lesssim 10^{22}$~cm$^{-2}$
unless either DLAs are detected or $T_s/f \gtrsim 500$~K on average.

Clearly a more productive approach to obtain the 
column density frequency distribution $f(N_{HI},X)$ at $z=0$ is via HI
21 cm emission lines \citep[e.g.,][]{zwaan05}, but 
beyond z$\sim0.2$, where the emission line flux density becomes
prohibitively weak for anything less than a square kilometer of
collecting area, one would use a distance-independent absorption line survey such as the
pilot survey presented here.  
Furthermore, ``blind'' 21~cm-based detections of DLAs can 
identify dusty sightlines that would otherwise be missed by optical/UV surveys, which is key to
identifying exceptionally rare molecular absorption systems 
and addressing the fraction of dusty DLAs.

\subsection{Future Surveys}\label{subsec:future}

As Figures \ref{fig:columns}, \ref{fig:fofN}, and \ref{fig:fofNrandom} demonstrate, even incremental improvements 
in survey sensitivity would lead to large improvements in column density sensitivity and statistics
because the source population grows as a power law as the 1.4~GHz flux density decreases 
(Figure \ref{fig:fluxes}).  But for strong sources, which provide probes of low column density, 
areal coverage remains important.

The first step should be to search the full ALFALFA survey for 21~cm
absorption lines; this pilot survey
covered 7\% of the full survey, so an order of magnitude increase in search path is possible using extant
data.  This will either produce detections of intervening absorption
line systems or lower the limits on the column 
density frequency distribution function, indicating a statistical
lower bound on spin temperature (modulo covering fraction).  It is already clear from Figure
\ref{fig:fofN} that the typical spin
temperature of HI gas at $z\sim0$ cannot be less than $T_s/f \simeq 100$~K.

Future facilities that could perform commensal or dedicated 21~cm absorption line surveys include
Square Kilometer Array (SKA) prototypes, including the 256-element 
Allen Telescope Array, the SKA Molonglo Prototype, the Australian SKA Pathfinder, and the Karoo Array 
Telescope (MeerKAT), as well as Epoch of Reionization (EoR) 
and Dark Ages telescopes, such as a lunar far-side array,
the Low Frequency Array (LOFAR), the Precision Array to Probe the EoR, and the Murchison Widefield Array.
These facilities will provide various combinations of areal coverage, bandwidth ($\delta z$ per source), 
and sensitivity.  As we have demonstrated here, sensitivity is key because it determines the number 
of sightlines available per solid angle of sky surveyed, and the power-law increase in radio source 
counts with decreasing flux density means that even slight improvements in sensitivity will lead to 
large gains in the size of absorption line surveys.  Bandwidth and areal coverage, in comparison, 
grow absorption line surveys linearly.

While absorption line detection is distance independent, which is an enormous advantage over 
redshifted emission line surveys, future HI absorption line surveys will have to contend with the issue of 
unknown redshifts of the continuum population and the loss of foreground sources in the total
redshift search path.

% [Comments on sensitivity and areal improvements required to make this really interesting...state
% rough parameters for an interesting HI study.]
% [Why bother with this when so much HI emission is detected in ALFALFA?  First, HI absorption detection is distance-independent, so future radio spectral line surveys below ~1100 MHz (z~xx)
% will lose sensitivity to HI emission, but will retain sensitivity to HI absorption.  Second, large fractional bandwidths translate into large spans in redshift, particularly in the sub-GHz regime, and the search and analysis techniques presented here will become more relevant and probe vast total redshift intervals...

\section{Conclusions}\label{sec:conclusions}

Our ``blind'' pilot survey for HI 21~cm absorption lines in a 517 deg$^2$ section of the ALFALFA survey
spanning $-650 ~{\rm km~s}^{-1}< cz < 17,500 ~{\rm km~s}^{-1}$ detected no new absorption line systems, 
but re-detected the intrinsic absorption in UGC~6081.  We calculate column density sensitivity limits toward 
a sample of 7296 radio sources and construct upper limit envelopes for the column density frequency 
distribution function $f(N_{HI},X)$ that are consistent with previous work at high and low redshift 
\citep[e.g.,][]{zwaan05,proch09,noterdaeme09}.  Our results suggest that higher redshift,
larger solid angle, larger bandwidth, and especially more sensitive surveys will be an 
astrophysically worthwhile pursuit in terms of constraining the 21~cm spin temperature and detecting 
dusty damped Ly$\alpha$ systems as well as DLAs toward optically faint but radio-loud illumination sources.
A subset of these HI 21~cm absorption systems will also host the 
exceptionally rare and elusive molecular absorption systems, which can be employed to 
study cold molecular gas at high redshift and to constrain the cosmic evolution of fundamental physical
constants.  

%\section{To Do}
%
% Address all red bits.
% 
% See if LUNAR issues/language can be worked in...
% 
% Check references both ways.
% 
% Get copy to RG and Burns for additions/feedback/vetting.
% 

\acknowledgements
The authors thank the many members of the ALFALFA team who have
contributed to the acquisition and processing of the ALFALFA
data set, especially Amelie Saintonge and Brian
Kent for the software they developed for general
implementation within in the ALFALFA data processing software package.
The authors thank Jason X. Prochaska for critical questions and help with the calculation 
of the column density frequency distribution function.  We also 
thank the anonymous referees for extremely helpful feedback.
RG and MPH are supported by NSF grant AST-0607007 and by
a grant from the Brinson Foundation.
The LUNAR consortium (http://lunar.colorado.edu), headquartered at
the University of Colorado, is funded by the NASA Lunar Science
Institute (via Cooperative Agreement NNA09DB30A) to investigate
concepts for astrophysical observatories on the Moon.
This research has made use of the NASA/IPAC Extragalactic Database (NED) 
which is operated by the Jet Propulsion Laboratory, California Institute of
Technology, under contract with NASA.  
This publication makes use of data products from the Two Micron All Sky Survey, which is a joint project of the University of Massachusetts and the Infrared Processing and Analysis Center/California Institute of Technology, funded by the National Aeronautics and Space Administration and the National Science Foundation.
Funding for the SDSS and SDSS-II has been provided by the Alfred P. Sloan Foundation, the Participating Institutions, the National Science Foundation, the U.S. Department of Energy, the National Aeronautics and Space Administration, the Japanese Monbukagakusho, the Max Planck Society, and the Higher Education Funding Council for England. The SDSS Web Site is http://www.sdss.org/.

{\it Facilities:}  \facility{Arecibo ()}, \facility{GBT ()}

\end{document}